# The rapidity and thermal motion induced transverse momentum distributions of identified charged particles produced in Au-Au collisions at RHIC energies

Zhi-Jin Jiang [1)], Jia-Qi Hui and Yu Zhang

*College of Science, University of Shanghai for Science and Technology, Shanghai 200093, China*

**Abstract** It is widely believed that the quark-gluon plasma (QGP) might be formed in the current heavy ion collisions. It is also widely recognized that the relativistic hydrodynamics is one of the best tools for describing the process of expansion and particlization of QGP. In this paper, by taking into account the effects of thermal motion, a hydrodynamic model including phase transition from QGP state to hadronic state is used to analyze the rapidity and transverse momentum distributions of identified charged particles produced in heavy ion collisions. A comparison is made between the theoretical results and experimental data. The theoretical model gives a good description to the corresponding measurements made in Au-Au collisions at RHIC energies.

## 1. Introduction

The primary goal of experimental program performed at Relativistic Heavy Ion Collider (RHIC) at Brookhaven National Laboratory (BNL) and at Large Hadron Collider (LHC) at CERN is to create the hot and dense matter consisting of partonic degrees of freedom, usually called the quark-gluon plasma (QGP), which is believed to have filled in the early universe several microseconds after the big bang. The calculations of Lattice Quantum Chromodynamics (LQCD) have predicted [1] that such matter may exists in the environment with critical temperature of about $T_c = 180\,\text{MeV}$ or energy density $\varepsilon_c = 2\,\text{GeV/fm}^3$. By means of the Bjorken estimation [2] and the measurements of PHENIX Collaboration at RHIC, the spatial energy density in central Au-Au collisions at $\sqrt{s_{\text{NN}}} = 200$ and 130 GeV is evaluated to be much higher than $\varepsilon_c$ [3]. Further studies have shown that QGP might be indeed generated in these collisions [4-7]. In fact,

---

it has long been argued that QGP might even have come into being in $p$-$p(\bar{p})$ collisions at the energies of Intersecting Storage Rings (ISR) and Super Proton Synchrotron (SPS) at CERN [8-12].

In the past decade, a number of bulk observables about charged particles, such as the Fourier harmonic coefficients $v_n$ of azimuth-angle distributions [13, 14], rapidity or pseudorapidity distributions [15-18] and transverse momentum distributions [19-24], have experienced a series of extensive investigations in heavy ion collisions at both RHIC and LHC energies. These investigations have offered us a compelling evidence that the matter created in heavy ion collisions exhibits a clear collective behavior, expanding nearly like an ideal fluid with very low viscosity. This sets up the prominent position of relativistic hydrodynamics in analyzing the properties of bulk observables in heavy ion collisions [25-42].

Apart from collective movement, the quanta of produced matter also have the components of thermal motion. The evolution of the produced matter is the convolution of the collective movement and thermal motion. To clarify the role of thermal motion in the expansions of the produced matter is the major subject of this paper. To this end, we may as usual ignore the minor collective flow in the transverse directions. The transverse movement of the produced matter is therefore only induced by the thermal motion.

The collective movement of produced matter in the longitudinal direction can be solved analytically. There are a few schemes in dealing with such precise calculations. In this paper, the hydrodynamic model proposed by N. Suzuki is employed [25]. Besides the analyticity, the other typical feature of this model is that it incorporates the effects of phase transition into solutions. This coincides with the current experimental observations as mentioned above. Hence, the employed model is more in line with the realistic situations. In addition, the model is related to the initial temperature of QGP, the sound speed in both partonic and hadronic media, the baryochemical potential, as well as the critical temperature of phase transition. This work may therefore help us understand various transport coefficients of expanding system.

In section 2, a brief introduction is given to the theoretical model [25], presenting its analytical formulations. The solutions are then used in section 3 to get the invariant multiplicity distributions of charged particles produced in heavy ion collisions which are in turn compared with the experimental measurements carried out by BRAHMS and PHENIX Collaboration in Au-Au collisions at RHIC energies of $\sqrt{s_{NN}}=200$ and 130 GeV [16-20]. The last section 4 is about conclusions.

## 2. A brief introduction to the model

The main content of the theoretical model [25] is as follows.

(1) In the process of expansions, the energy and momentum of fluid are conserved. Hence, the movement of fluid follows the continuity equation

$$\frac{\partial T^{\mu\nu}}{\partial x^{\nu}}=0, \quad \mu,\nu=0,1, \tag{1}$$

where $x^{\nu}=(x^0, x^1)=(t, z)$, $t$ is the time and $z$ is the longitudinal coordinate along beam direction.

$T^{\mu\nu}$ is the energy-momentum tensor, which, for an ideal fluid, takes the form

$$T^{\mu\nu}=(\varepsilon+p)u^{\mu}u^{\nu}-pg^{\mu\nu}, \tag{2}$$

where $g^{\mu\nu}=g_{\mu\nu}=\mathrm{diag}(1,-1)$ is the metric tensor.

$$u^{\mu}=(u^0, u^1)=(\cosh y_{\mathrm{L}}, \sinh y_{\mathrm{L}}), \quad u^{\mu}u_{\mu}=1, \tag{3}$$

is the 4-velocity of fluid, $y_{\mathrm{L}}$ is its rapidity. $\varepsilon$ and $p$ in Eq. (2) are the energy density and pressure of fluid, which meet the thermodynamical relations

$$\varepsilon+p=Ts, \quad \mathrm{d}\varepsilon=T\mathrm{d}s, \quad \mathrm{d}p=s\mathrm{d}T, \tag{4}$$

where $T$ and $s$ are the temperature and entropy density of fluid, respectively. To fulfill Eq. (1), another relation, namely the equation of state

$$\frac{\mathrm{d}p}{\mathrm{d}\varepsilon}=\frac{s\mathrm{d}T}{T\mathrm{d}s}=c_{\mathrm{s}}^2 \tag{5}$$

is needed, where $c_{\mathrm{s}}$ is the sound speed of fluid, which takes different values in partonic and in hadronic phase.

(2) In order to solve Eq. (1), Khalatnikov scheme is used. To this end, Eq. (1) is separately projected to the direction of $u_{\mu}$ and the direction perpendicular to $u_{\mu}$. This leads to two equations

$$\frac{\partial(su^\nu)}{\partial x^\nu}=0, \tag{6}$$

$$\frac{\partial(T\sinh y_L)}{\partial t}+\frac{\partial(T\cosh y_L)}{\partial z}=0. \tag{7}$$

Eq. (6) is the continuity equation for entropy conservation. Eq. (7) means the existence of a scalar function $\phi$ satisfying relations

$$\frac{\partial\phi}{\partial t}=T\cosh y_L,\ \frac{\partial\phi}{\partial z}=-T\sinh y_L. \tag{8}$$

From $\phi$ and Legendre transformation, Khalatnikov potential $\chi$ is introduced *via*

$$\chi=\phi-tT\cosh y_L+zT\sinh y_L. \tag{9}$$

In terms of $\chi$, the variables $t$ and $z$ can be expressed as

$$\begin{aligned}t&=\frac{e^\theta}{T_0}\left(\frac{\partial\chi}{\partial\theta}\cosh y_L+\frac{\partial\chi}{\partial y_L}\sinh y_L\right),\\ z&=\frac{e^\theta}{T_0}\left(\frac{\partial\chi}{\partial\theta}\sinh y_L+\frac{\partial\chi}{\partial y_L}\cosh y_L\right),\end{aligned} \tag{10}$$

where $T_0$ is the initial temperature of fluid and $\theta=\ln(T_0/T)$. Hence, through above equation, the coordinate base of $(t, z)$ is transformed to that of $(\theta, y_L)$. Eq. (6) is translated to the so called telegraphy equation

$$\frac{\partial^2\chi}{\partial\theta^2}-2\beta\frac{\partial\chi}{\partial\theta}-\frac{1}{c_s^2}\frac{\partial^2\chi}{\partial y_L^2}=0,\ \beta=\frac{1-c_s^2}{2c_s^2}. \tag{11}$$

(3) Along with the expansions of matter created in collisions, its temperature becomes lower and lower. As the temperature drops from the initial $T_0$ to the critical $T_c$, phase transition occurs. The matter transforms from QGP state to hadronic state. The produced hadrons are initially in the violent and frequent collisions. The major part of these collisions is inelastic. Hence, the abundances of identified hadrons are in changing. Furthermore, the mean free paths of these primary hadrons are very short. The movement of them is still like that of a fluid meeting Eq. (11) with only difference being the value of $c_s$. In QGP, $c_s=c_0=1/\sqrt{3}$, which is the sound speed of a massless ideal fluid, being the maximum of $c_s$. In the hadronic state, $c_s=c_h<c_0$. At the point of phase transition, that is as $T=T_c$, $c_s$ is discontinuous.

(4) The solution of Eq. (11) for the sector of QGP is [25]

$$\chi_0(\theta, y_L) = \frac{q_0 c_0}{2} e^{\beta_0 \theta} I_0\left(\beta_0 c_0 \sqrt{y_0^2(\theta) - y_L^2}\right), \tag{12}$$

where $q_0$ is a constant determined by tuning the theoretical results to experimental data. $I_0$ is the 0th order modified Bessel function of the first kind, and

$$\beta_0 = \frac{1 - c_0^2}{2 c_0^2} = 1, \quad y_0(\theta) = \frac{\theta}{c_0}. \tag{13}$$

In the sector of hadrons, the solution of Eq. (11) is [25]

$$\chi_h(\theta, y_L) = \frac{q_0 c_0}{2} B(\theta) I_0\left[\lambda(\theta, y_L)\right], \tag{14}$$

where

$$B(\theta) = e^{\beta_h(\theta - \theta_c) + \beta_0 \theta_c}, \quad \lambda(\theta, y_L) = \beta_h c_h \sqrt{y_h^2(\theta) - y_L^2},$$
$$\beta_h = \frac{1 - c_h^2}{2 c_h^2}, \quad y_h(\theta) = \frac{\theta - \theta_c}{c_h} + \frac{\theta_c}{c_0}, \quad \theta_c = \ln\left(\frac{T_0}{T_c}\right). \tag{15}$$

It is evident that, if $c_h = c_0$, then $\beta_h = \beta_0$, $B(\theta) = e^{\beta_0 \theta}$, $y_h = y_0$, and thus $\chi_h = \chi_0$. At the point of phase transition, $T = T_c$, $\theta = \theta_c$, $B(\theta) = e^{\beta_0 \theta_c}$, $y_h = y_0 = \theta_c / c_0$. Then

$$\chi_0(\theta, y_L)\big|_{T=T_c} = \frac{q_0 c_0}{2} e^{\beta_0 \theta_c} I_0\left(\beta_0 \sqrt{\theta_c^2 - c_0^2 y_L^2}\right),$$
$$\chi_h(\theta, y_L)\big|_{T=T_c} = \frac{q_0 c_0}{2} e^{\beta_0 \theta_c} I_0\left(\frac{\beta_h c_h}{c_0} \sqrt{\theta_c^2 - c_0^2 y_L^2}\right), \tag{16}$$

$\chi_h(\theta, y_L)\big|_{T=T_c} \neq \chi_0(\theta, y_L)\big|_{T=T_c}$. That is, the potential $\chi$ is discontinuous at point of $T = T_c$.

## 3. The rapidity and thermal motion induced transverse momentum distributions of identified charged particles

With the expansions of hadronic matter, its temperature continues becoming lower. According to the prescription of Cooper-Frye [43], as the temperature drops to the so called chemical freeze-out temperature $T_{FO}$, the inelastic collisions among hadrons cease. The abundances of identified hadrons maintain unchanged becoming the measured results in experiments. The invariant multiplicity distributions of charged particles are given by [25, 43, 44]

$$\frac{d^2 N}{2\pi p_T dy dp_T} = \frac{q_0 c_0}{2} \frac{A(b)}{(2\pi)^3} \int \left(\cosh y \frac{dz}{dy_L} - \sinh y \frac{dt}{dy_L}\right)\bigg|_{T=T_{FO}}$$

$$\times \frac{m_T \cosh(y-y_L)}{\exp\{[m_T \cosh(y-y_L)-\mu_B]/T_{FO}\}+\delta} dy_L, \tag{17}$$

where $A(b)$ is the area of overlap region of collisions, being the function of impact parameter $b$ or centrality cuts. $m_T = \sqrt{m^2+p_T^2}$ is the transverse mass of produced charged particle with rest mass $m$. $\mu_B$ in Eq. (17) is the baryochemical potential. For Fermi charged particles, $\delta=1$ in the denominator in Eq. (17), and for Bosons, $\delta=-1$. The meaning of Eq. (17) is evident. The part of integrand in the round brackets is proportional to the rapidity density of fluid resulted from the collective movement along longitudinal direction [43]. The rest part is the energy of the charged particles in the state with temperature $T_{FO}$ resulted from thermal motion. Eq. (17) is just the convolution of these two parts.

From the relations of Eq. (10), it can be shown that

$$\left(\cosh y \frac{dz}{dy_L} - \sinh y \frac{dt}{dy_L}\right)\bigg|_{T=T_{FO}}$$

$$= \left[-\frac{1}{T}\frac{\partial}{\partial y_L}\left(\chi_h + \frac{\partial \chi_h}{\partial \theta}\right)\sinh(y-y_L) + \frac{1}{T}c_h^2 \frac{\partial}{\partial \theta}\left(\chi_h + \frac{\partial \chi_h}{\partial \theta}\right)\cosh(y-y_L)\right]\bigg|_{T=T_{FO}}.$$

Substituting $\chi_h$ of Eq. (14) into above equation, it becomes

$$\left(\cosh y \frac{dz}{dy_L} - \sinh y \frac{dt}{dy_L}\right)\bigg|_{T=T_{FO}}$$

$$= \frac{1}{T_{FO}}(\beta_h c_h)^2 B(\theta_{FO})\left[S(\theta_{FO}, y_L)\sinh(y-y_L) + C(\theta_{FO}, y_L)\cosh(y-y_L)\right], \tag{18}$$

where

$$S(\theta_{FO}, y_L) = \frac{\beta_h y_L}{\lambda(\theta_{FO}, y_L)}\left\{\frac{\beta_h c_h y_h(\theta_{FO})}{\lambda(\theta_{FO}, y_L)}I_0[\lambda(\theta_{FO}, y_L)]\right.$$

$$\left. + \left[\frac{\beta_h+1}{\beta_h} - \frac{2\beta_h c_h y_h(\theta_{FO})}{\lambda^2(\theta_{FO}, y_L)}\right]I_1[\lambda(\theta_{FO}, y_L)]\right\}, \tag{19}$$

$$C(\theta_{FO}, y_L) = \left\{\frac{\beta_h+1}{\beta_h} + \frac{[\beta_h c_h y_h(\theta_{FO})]^2}{\lambda^2(\theta_{FO}, y_L)}\right\}I_0[\lambda(\theta_{FO}, y_L)]$$

$$+\frac{1}{\lambda(\theta_{FO}, y_L)}\left\{\frac{y_h(\theta_{FO})}{c_h}+1-\frac{2[\beta_h c_h y_h(\theta_{FO})]^2}{\lambda^2(\theta_{FO}, y_L)}\right\} I_1[\lambda(\theta_{FO}, y_L)], \qquad (20)$$

where $I_1$ is the 1st order modified Bessel function of the first kind.

The integral interval of $y_L$ in Eq. (17) is $[-y_h, y_h]$. By applying Eqs. (17)-(20), together with the definitions in Eq. (15), we can get the rapidity and thermal motion induced transverse momentum distributions of identified charged particles as shown Figs. 1, 2 and 3.

Fig. 1 shows the rapidity distributions of identified charged particles in central Au-Au collisions at $\sqrt{s_{NN}} = 200$ GeV. The solid dots are the experimental measurements [16-18]. The solid curves are the results from Eq. (17). It can be seen that, except for proton $p$, the theoretical predictions are in good agreement with experimental data. For proton $p$, the experimental data appear an evident uplift in the rapidity interval between $y = 2.0$ and $3.0$. This may be resulted from parts of leading particles [37, 38], which are free from the description of hydrodynamics.

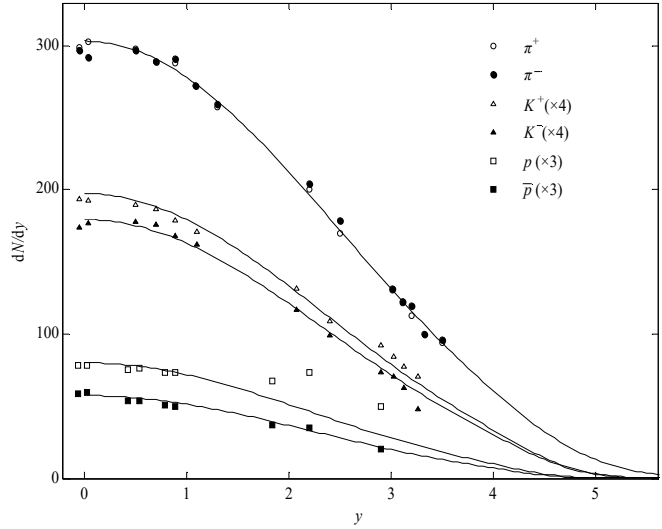

Fig. 1. The rapidity distributions of $\pi^\pm$, $K^\pm$ and $p(\bar{p})$ in central Au-Au collisions at $\sqrt{s_{NN}} = 200$ GeV. The solid dots are the experimental measurements [16-18]. The solid curves are the results from Eq. (17).

Figs. 2 and 3 are for centrality dependences of thermal motion induced transverse momentum distributions of $\pi^\pm$, $K^\pm$ and $p(\bar{p})$ in Au-Au collisions at $\sqrt{s_{NN}} = 200$ and 130 GeV, respectively. The left columns are for positive charged particles. The right columns are for negative ones. The centrality cuts are 0-5%, 5-10%, 10-15%, 15-20%, 20-30%, 30-40%, 40-50%,

50-60%, 60-70%, 70-80%, 80-92% and 60-92%, counted from top to bottom in Fig. 2. The centrality cuts in Fig. 3 are 0-5%, 5-15%, 15-30%, 30-60% and 60-92% counted from top to bottom. The solid dots are the experimental measurements [19, 20]. The solid curves are the hydrodynamic predictions of Eq. (17). It can be seen that the theoretical results are in good accordance with experimental data for transverse momentum up to about $p_T = 1.1 \text{GeV/c}$.

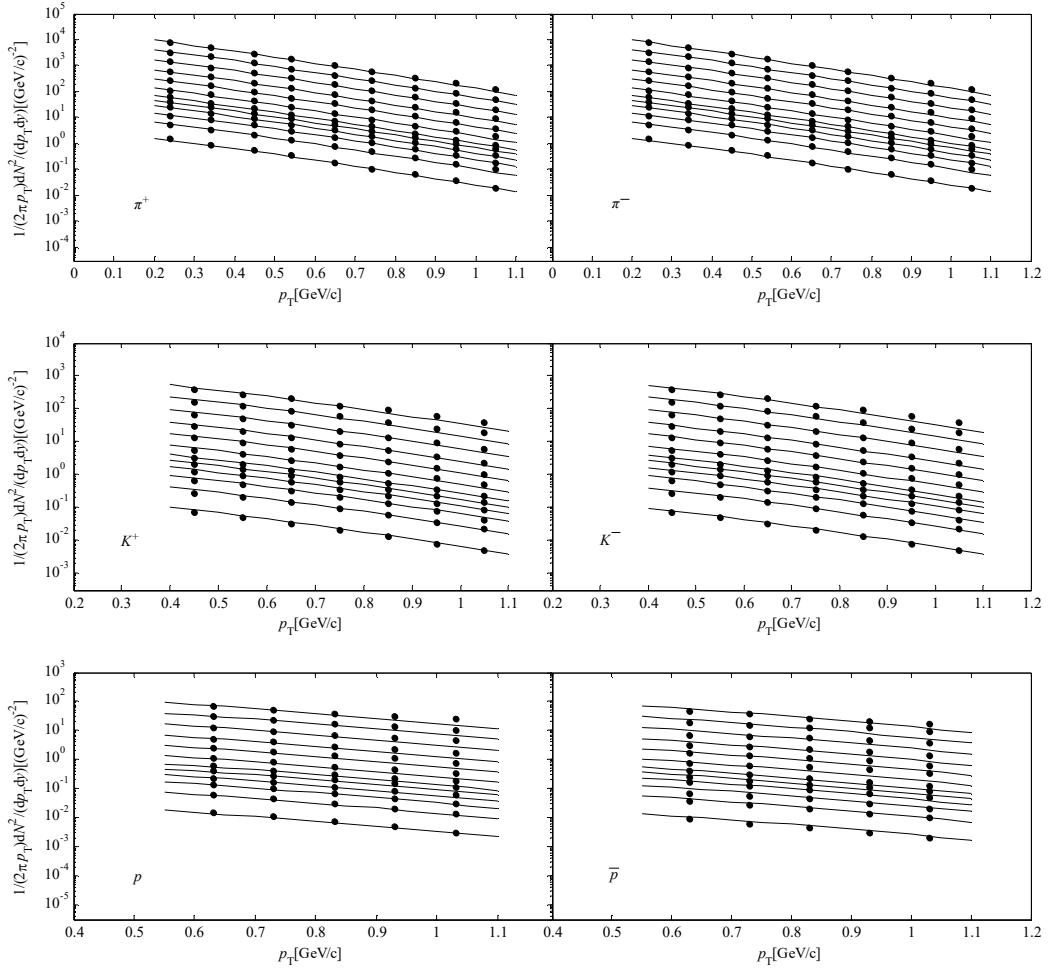

Fig. 2. The centrality dependences of transverse momentum distributions of identified charged particles produced at mid-rapidity of $|y| \leq 1$ in Au-Au collisions at $\sqrt{s_{NN}} = 200$ GeV. The solid dots are the experimental data [19]. The solid curves are the hydrodynamic results of Eq. (17). The centrality cuts counted from top to bottom in each panel are 0-5%, 5-10%, 10-15%, 15-20%, 20-30%, 30-40%, 40-50%, 50-60%, 60-70%, 70-80%, 80-92% and 60-92%, respectively.

In calculations, $T_c$ in Eq. (15) takes the well-recognized value of $T_c = 180 \text{MeV}$. $c_h$ in Eq. (15) takes the value of $c_h = 0.35$ from the investigations of Refs. [44-47]. The chemical freeze-out temperature $T_{FO}$ in Eqs. (19) and (20) takes a constant of $T_{FO} = 160 \text{MeV}$ from the studies of Ref. [23], which also show that the baryochemical potential $\mu_B$ in Eq. (17) is about equal to 20 and 27 MeV in Au-Au collisions at $\sqrt{s_{NN}}$ =200 and 130 GeV, respectively. The initial temperature $T_0$ in Eq. (15) takes the values of $T_0 = 0.70$ and 0.55 GeV referring to those given in Ref. [44] in central $\sqrt{s_{NN}} = 200$ and 130 GeV Au-Au collisions. This allows us to determine the constant $q_0$ in Eq. (17) to be

$$q_0\left(\pi^+/\pi^-;\ K^+/K^-;\ p/\bar{p}\right) = 2.50/2.50;\ 1.20/1.16;\ 2.43/1.82\ (200\,\text{GeV}),$$
$$q_0\left(\pi^+/\pi^-;\ K^+/K^-;\ p/\bar{p}\right) = 3.50/3.50;\ 2.02/1.87;\ 9.78/7.02\ (130\,\text{GeV}).$$
(21)

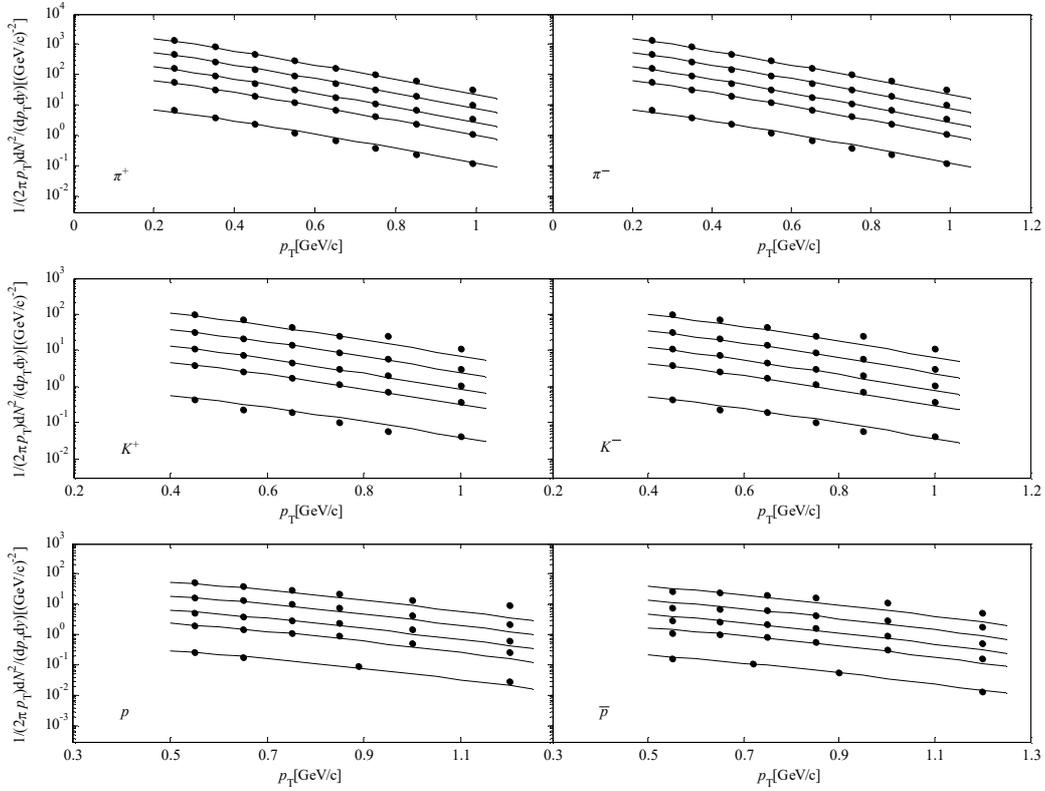

Fig. 3. The centrality dependences of transverse momentum distributions of identified charged particles produced at mid-rapidity of $|y| \leq 1$ in Au-Au collisions at $\sqrt{s_{NN}} = 130$ GeV. The solid dots are the experimental data [20]. The solid curves are the hydrodynamic results of Eq. (17). The centrality cuts counted from top to bottom in each panel are 0-5%, 5-15%, 15-30%, 30-60% and 60-92%, respectively.

The fitted values of Eq. (21) give the ratios

$$q_0\left(\pi^+\right)/q_0\left(\pi^-\right) = 1,\ q_0\left(K^+\right)/q_0\left(K^-\right) = 1.03,\ q_0(p)/q_0(\bar{p}) = 1.34\ (200\,\text{GeV}),$$
$$q_0\left(\pi^+\right)/q_0\left(\pi^-\right) = 1,\ q_0\left(K^+\right)/q_0\left(K^-\right) = 1.08,\ q_0(p)/q_0(\bar{p}) = 1.39\ (130\,\text{GeV}).$$
(22)

These ratios are well consistent with the relative abundances of particles and antiparticles presented in Ref. [23]. This consistence may be attributed to the fact that, in case of adopting a common $T_{FO}$ for different charged particles, the integral part of Eq. (17) is the same for particles and antiparticles. Hence, $q_0$ should be proportional to the abundance of the corresponding particles.

Keeping the constants $q_0$ in Eq. (21) unchanged, we can get the fitted $T_0$ falling in the ranges of $(0.70 \sim 0.33\,\text{GeV})$ and $(0.55 \sim 0.26\,\text{GeV})$ from small to large centrality cuts for $\sqrt{s_{NN}} = 200$ and 130 GeV, respectively. As can be expected, $T_0$ decreases with increasing centrality cuts.

## 4. Conclusions

In order to see the importance of thermal motion in the expansions of the hot and dense matter created in high energy heavy ion collisions, the minor transverse collective flow is as usual ignored. The movement of matter in the transverse directions is therefore only induced by the thermal motion. Convoluted with longitudinal collective flow, we can get the invariant multiplicity distributions of charged particles.

The longitudinal collective flow follows the relativistic hydrodynamics. The most striking feature of the hydrodynamic model used in this paper is that it takes the phase transition into account, which might have appeared in the current experiments in nucleus or even in hadron collisions. Hence, the adopted model might reflect a more realistic process of spatiotemporal evolution of matter created in collisions.

The model contains a rich information about the transport coefficients of fluid, such as the sound speed in QGP $c_0$, the sound speed in hadronic phase $c_h$, the phase transition temperature $T_c$, the chemical freeze-out temperature $T_{FO}$, the baryochemical potential $\mu_B$ and the initial temperature $T_0$. With the exception of $T_0$, the other five coefficients take the values either from the well-known theoretical results or from experimental measurements. As for $T_0$, there are no widely accepted results so far. In our calculations, $T_0$ in the most central Au-Au collisions at both $\sqrt{s_{NN}}=200$ and 130 GeV takes the values from other researches, and in the rest centrality cuts, $T_0$ is determined by maintaining $q_0$ unchanged and comparing the theoretical results with experimental data. The ratios of $q_0$ for positive and negative particles determined in this paper are in good accordance with current experimental measurements.

The theoretical model can give a good description to the rapidity distributions measured in central Au-Au collisions at $\sqrt{s_{NN}}=200$ GeV. For the transverse momentum distributions measured in Au-Au collisions at $\sqrt{s_{NN}}=200$ and 130 GeV, the theoretical results match up well with experimental data for the transverse momentum up to about $p_T=1.1\,\text{GeV}/c$. Certainly, the transverse collective flow is objectively exist even though it is minor. It should affect the transverse motion of the matter. The existing investigations have shown that if the minor transverse collective flow is taken into account, the describable region is extended to about $p_T<2.0\,\text{GeV}/c$ [20-22]. Furthermore, to describe the experimental data in the region of $p_T\geq 2.0\,\text{GeV}/c$, some other mechanisms, such as the method of blast-wave and perturbative QCD [24], should be introduced. Considering that the overwhelming majority of charged particles situates in the region of $p_T<1.1\,\text{GeV}/c$, we may conclude that the thermal motion plays a non-ignorable role in the transverse expansions of the matter produced in high energy heavy ion collisions.

## Acknowledgments

This work is supported by the Shanghai Key Lab of Modern Optical System.